\documentstyle[11pt,aaspp4]{article}

\newcommand{\fun}{\hbox{\ erg cm$^{-2}$ s$^{-1}$} }
\newcommand{\lun}{\hbox{\ erg s$^{-1}$} }

\slugcomment{To be published in {\it ApJ Letter}}

\lefthead{Rosati et al.}
\righthead{ROSAT Deep Cluster Survey}

\begin{document}

\title{The ROSAT Deep Cluster Survey: 
	the X-ray Luminosity Function out to $z=0.8$}

\author{Piero Rosati,\altaffilmark{1,2,3,4} 
        Roberto Della Ceca,\altaffilmark{5}
	Colin Norman,\altaffilmark{2}
	and Riccardo Giacconi\altaffilmark{1} }
	

\altaffiltext{1}{ESO - European Southern Observatory, D-85748 Garching b. M\"unchen, Germany; prosati@eso.org} 
\altaffiltext{2}{Department of Physics \& Astronomy, The Johns Hopkins University, Baltimore, MD, 21218}
\altaffiltext{3}{Visiting Astronomer, Kitt Peak National Observatory. 
KPNO is operated by AURA, Inc.\ under contract to the National Science
Foundation.} 
\altaffiltext{4}{Visiting Astronomer, Cerro Tololo Inter-American Observatory. 
CTIO is operated by AURA, Inc.\ under contract to the National Science
Foundation.} 
\altaffiltext{5}{Osservatorio Astronomico di Brera, 20121 Milano, Italy}


\begin{abstract}
We present the X-ray Luminosity Function (XLF) of the ROSAT Deep
Cluster Survey (RDCS) sample over the redshift range 0.05--0.8.  Our
results are derived from a complete flux-limited subsample of 70 galaxy
clusters, representing the brightest half of the total sample, which have been
spectroscopically identified down to the flux limit of
$4\times10^{-14}$ \fun (0.5--2.0 keV) and have been selected
via a serendipitous search in ROSAT-PSPC pointed observations.
The redshift baseline is large enough that evolutionary effects can be
studied within the sample. The local XLF ($z\le 0.25$) is found to be in
excellent agreement with previous determinations using the ROSAT All-Sky 
Survey data.
The XLF at higher redshifts, when combined with the deepest number counts constructed
to date ($f>2\times10^{-14} \fun$), reveal no significant evolution at
least out to $z=0.8$, over a luminosity range $2\times 10^{42}-3\times 10^{44}$
\lun in the [0.5-2 keV] band.
These findings extend the study of cluster evolution to the highest
redshifts and the faintest fluxes probed so far in X-ray surveys. They
complement and do not necessarily conflict with those of the Einstein
Extended Medium Sensitivity Survey, leaving the possibility of negative
evolution of the brightest end of the XLF at high redshifts.
\end{abstract}



\section{Introduction}
Large homogeneous samples of galaxy clusters, spanning a wide range of
redshifts, have long been considered powerful tools to study the
evolution of the large scale structures in the Universe. Unfortunately,
finding clusters at cosmologically interesting lookback times, i.e.
$z>0.5$, not to mention defining a complete sample, is a time consuming
and difficult task. As a result, the much-needed observational
constraints for theories of cluster formation have not been
forthcoming.

In an effort to remedy this situation, we have embarked on a project, the
ROSAT Deep Cluster Survey (RDCS), aimed at constructing a large
homogeneous sample of distant galaxy clusters selected solely on the
basis of their X-ray properties  from ROSAT-PSPC
pointed observations. Initial results have been presented in \cite{R95} (R95).
Pure X-ray selection leads to a selection function which can
be modelled in a relatively straightforward way, being essentially that
of a flux limited sample. As a result, distribution functions of
observables, such as the number counts,  the X-ray Luminosity
Function (XLF) and the redshift distribution, can be directly
compared with theories of structure formation.
Until very recently, the only available sample of X-ray selected clusters
 at high redshifts was the one compiled from the Einstein Extended
Medium Sensitivity Survey (EMSS) (\cite{gio90}, \cite{hen92} (H92)).
From a complete subsample of 67 clusters at  $0.14<z<0.6$, the authors found
evidence of a statistically significant negative evolution of the XLF
at $z\gtrsim 0.3$ and 
$L_{X \rm{\footnotesize{[0.3-3.5]~keV}}}  \gtrsim 5\times 10^{44} \lun$. 
Earlier claims 
of strong evolution at lower redshifts
(\cite{edg90}) have been ruled out by a large compilation of X-ray
clusters from the ROSAT All-Sky Survey (RASS) (\cite{ebe97}). More
recently, the EMSS findings have been challenged by a re-analysis of the
same sample supplemented by ROSAT data (\cite{nic97}).  Covering a very
large solid angle (735 deg$^2$), but being relatively shallow, the EMSS sample can probe mostly the bright end
of the XLF and can provide weak constraints on the evolution of the
XLF at higher redshifts since only 6 EMSS clusters lie at $z > 0.5$.
On the other hand,  the RDCS reaches considerably fainter fluxes, but covers a
much smaller solid angle and therefore probes a complementary region 
in the $L_X-z$ plane  to the EMSS.
Surveys similar to the RDCS, utilizing the PSPC archival data and
somewhat different selection techniques, are now well underway 
(RIXOS (\cite{cas95}), SHARC
(\cite{col97}, \cite{bur97}), WARPS (\cite{sha97})).

In this Letter, we present the  XLF constructed from a
flux-limited, spectroscopically confirmed subsample of the RDCS. This subsample
is already as large as the EMSS  sample and covers a wider
redshift range, from the local universe (z=0.05) out to $z\simeq
0.8$. Therefore, we can compare our low-redshift luminosity function with
recent determinations from cluster samples compiled from the ROSAT All-Sky
Survey (\cite{B96}, \cite{ebe97}). 
Furthermore, we can compare, for the first time, the space density of high
redshift clusters with the findings of the EMSS and extend the study of
cluster evolution to even higher redshifts.
We adopt $H_0 = 50$ km s$^{-1}$ Mpc$^{-1}$ and $q_0 = 0.5$  througout.

\section{The Cluster Sample}
A full discussion of  
the analysis of the X-ray data and the selection technique for the RDCS
sample is presented  in R95. Cluster candidates are selected from
a serendipitous search for extended X-ray sources in deep pointed
observations drawn from the ROSAT-PSPC archive. A wavelet-based
technique, which is particularly efficient at reducing the effects of
confusion and is not biased against low surface
brightness features, is used to detect and characterize X-ray sources. 
A control sample of about 5000 point sources is used to establish the mean
and the variance of the PSF at different off-axis angles ($\theta$),
thereby assessing the statistical significance for a source to be
extended. Cluster candidates are selected as those sources
extended at 99\% confidence level and having a flux $f_{-14}=f_{X}
[0.5-2.0]\hbox{keV}/(10^{-14}\fun) >1$ and $\theta < 18$ arcmin.  This
selection technique yielded 160 cluster candidates over an area of
about 48 deg$^2$, drawn from 170 X-ray fields scattered across the two
galactic caps ($|b|>20^\circ$).

The completeness flux limit of the survey is determined by the flux
level at which extended and point-like emission can be reliably
distinguished. In addition to the source flux, this critically depends on the
off-axis angle $\theta$ within which the candidates are selected, due
to PSF degradation. From a consideration of surface brightness dimming
alone, one expects the fraction of clusters which are
unresolved by the PSPC at high
redshifts to increase at faint flux levels and at high off-axis angles.
These effects are analyzed in detail elsewhere (Rosati \& Della Ceca,
in preparation). The selection function is modelled and quantified
using a combination of simulations, control samples of known distant
clusters and by studying cluster number counts as a function of
$(f_X,\theta)$.  This analysis allows two statistically complete
and independent samples to be defined: one deeper
(Sample A), covering $\sim\! 33$ deg$^2$ ($f_{-14} >1, \theta \le
15\arcmin$) and one shallower (Sample B) with $f_{-14} >6$ and $15\arcmin <
\theta\le 18\arcmin$. By combining these distinct samples the
surveyed area increases to $\sim\! 48$ deg$^2$.
We have chosen a higher flux limit for Sample B since it becomes
significantly incomplete for $f_{-14} < 5$.  The incompleteness of
Sample A is measured to be negligible for $f_{-14} >4$, is about 10\%
at $f_{-14}\simeq 2$ and drops to $\sim\! 20\!-\!25\%$  at $1< f_{-14} <2$.

 The sky coverage $\Omega$ of the survey, as emphasized in R95,
also depends on the intrinsic angular size $(\Theta$) of the
source for a given flux, i.e. $\Omega = \Omega (f,\Theta)$. Here, for
the sake of simplicity, we have reduced this bivariate function to a
function of the flux only, by choosing a value $\bar{\Theta}$ equal to
the median angular size  of the clusters in the whole sample after
deconvolving the PSF ($\bar{\Theta}\simeq 58\arcsec$).
We have verified that the use of this effective sky coverage is
equivalent to the general method described in R95. The function $\Omega
= \Omega (f,\bar{\Theta})$ for Sample A and its extension ($A+B$) is
shown in fig.1.

\section{Optical Follow-up Observations}

In order to identify these cluster candidates, we have
undertaken a large optical follow-up program, consisting of deep
imaging using the KPNO 4m and 2.1m and the CTIO 4m and 1.5m, and
multislit spectroscopy carried out with KPNO 4m and the ESO 3.6m. 
The imaging survey in V and I bands is now 90\% complete and has
revealed a very high success rate of identification, with about 115
new clusters (or groups) confirmed to date, i.e. displaying a significant
overdensity of galaxies around the peak of the X-ray emission.  More
importantly, the spectroscopic follow-up work has secured 95 cluster
redshifts so far, spanning the range 0.045--0.83.  Most of the
spectroscopic identifications are based on more than 3 redshifts within
1500 km/s, with a median value of 5 members per cluster.  
A significant fraction of the newly discovered clusters lie at high
redshift, about one-third at $z>0.4$ and a quarter at $z>0.5$, making the
RDCS the largest sample of spectroscopically confirmed distant clusters
compiled to date. 
Details on the optical observations of the whole RDCS
catalogue will be published elsewhere.

\section{The X-ray Luminosity Function}

The imaging and spectroscopic identification of the cluster candidates 
in Sample A is 97\% complete down to $f_{-14} = 4$, hence 
a complete flux limited subsample of 70 clusters with measured
redshifts can be defined\footnote{
 To date, only two candidates have not been
identified down to the flux limit of $f_{-14}=4$. These are extended X-ray
sources which may be either a blend of multiple sources, for which we did not
find an obvious optical counterpart, or high redshift clusters, for
which deeper imaging is required.}.
We split the sample in three redshift shells, [0.045--0.25], [0.25--0.5], 
[0.5-0.85], where we have respectively 33, 23, 14 clusters.
The luminosities in the observed [0.5--2.0] keV band range from
$3.5\times 10^{44}$ to $1\times 10^{42}\lun$ and thus probe the region
of moderately rich clusters to poor groups.  A non-parametric
representation of the XLF has been obtained using the $1/V_{a}$ method
of Avni and Bahcall (1980).  The application of this procedure for
deriving the XLF in different redshift shells is fully described in
Maccacaro et al., 1991.  For each luminosity bin containing {\it n}
objects, the differential XLF is computed as,
$dN(L)/dL = {\sum^{n}_{i=1} 1/(V_{S_{i}} \Delta L) }$,
where $V_{S_{i}}$ represents the total search volume for object $i$  
 and $\Delta L$ is the width of
the luminosity bin.  Bins of equal logarithmic width
$\Delta Log\ L = 0.3$ have been used.
The corresponding $68\%$ error bars have been determined  
 using Poissonian statistics (following Wolter et al. 1994). 
A power law spectrum with energy index 0.5, which well approximates
a Raymond-Smith spectrum over a large temperature range in the 
[0.5-2.0] keV band, has been used for computing  k-corrections.
Count rates have been converted to fluxes according to R95.
 
The XLF derived is shown in fig.2. We point out the
excellent agreement between the local XLF of the RDCS, i.e. in the
lowest redshift bin, and two independent determinations of the local
XLF from  \cite{ebe97} and \cite{B96} (B96). Both these surveys use
RASS data but completely different selection
techniques.  The Bright Cluster Sample (BCS) of \cite{ebe97} is an
X-ray flux limited sample out to $z=0.3$ thus covering the intermediate
and bright end of the XLF, whereas the B96 sample is an optically
selected, volume complete, sample of nearby groups and poor clusters
($z\le 0.15$) which probes mostly the faint end, down to luminosities
well below $10^{42} \lun$.
The RDCS XLF in the first redshift shell is well fit by
a power law in the form $n(L_{44})= KL_{44}^{-\alpha}$, where $L_{44}$ is
the X-ray luminosity in the the rest frame band [0.5-2.0]keV in units of $10^{44}\lun$
and K is in units of $10^{-7}\hbox{Mpc}^{-3} [L_{44}]^{(\alpha-1)}$. 
A maximum-likelihood fit yields:  $\alpha= 1.83\pm 0.15$ and $K=3.26\pm 0.57$.
These parameters compare favorably with those of B96, 
 $\alpha= 1.71\pm 0.19$ and $K=3.50\pm 0.77$ (after rescaling
to $H_0=50$ km s$^{-1}$ Mpc$^{-1}$), and the power law part of the Schechter 
XLF of the BCS sample, $\alpha= 1.85\pm 0.09$ and
$K=3.32(^{+0.36}_{-0.33})$.
A different cluster sample, analogous to the BCS, compiled from the RASS
in southern sky provides similar results (\cite{sab96}).
The excellent agreement found between these independent determinations is
 by no means trivial since the RDCS is the only
sample of the four whose selection {\sl is not driven by any optical
information}, but purely by the X-ray properties of clusters. Indeed,
it suggests that systematic effects in local cluster samples
seem to be now well under control.

The inspection of the RDCS XLF in the higher redshift bins does not
show any significant evolution, at least out to $z\simeq 0.8$, over the
probed luminosity range. 
Based on the number of high redshift clusters in the SHARC-South sample
\cite{col97} also found no evidence of significant negative evolution. 
More recently, \cite{bur97} have contructed the XLF from the same 
small SHARC-S sample, 16 clusters in the range $0.3<z<0.7$,
and have reached similar conclusions to the ones reported in this work. 
The large fraction of high redshift clusters in the
RDCS cannot be reconciled with the results of the RIXOS survey (\cite{cas95}).
A factor of 4 more RDCS clusters have been identified at $z>0.4$
compared to the RIXOS survey down to the same flux limit ($f_{-14}=3$)
and over the same solid angle.  This difference may partly be due to
incompleteness or small number statistics in RIXOS, but also may
reflect an overestimate of the sky coverage of the RIXOS survey at low
fluxes ($\Omega(f,\Theta)$ is assumed constant) which could mimic a
dearth of clusters at high redshifts (\cite{rdc96}).

Given its relatively small surveyed area,
the RDCS cannot probe the XLF {\sl above} $L_X\simeq 3\times 10^{44} \lun$.
This is the luminosity range where the EMSS sample finds evidence of
negative evolution, i.e. a steepening of the high end of the XLF at
$z\gtrsim 0.3$ (\cite{gio90}). 
A direct comparison of the cluster volume densities of the RDCS and EMSS 
samples is illustrated in fig.3, where we recompute the XLF of the RDCS sample
in the redshift shell [0.3--0.6] (24 objects), the highest redshift shell of 
the EMSS sample.
The XLF data points of the EMSS sample (open and filled squares) have
been converted to the 0.5--2.0 keV band by recomputing the XLF using
the sample, the sky coverage and the method of H92, updating a few redshifts 
and identifications according to \cite{giolup94}.
A power law  spectral model  with energy index 0.5 has been used to convert fluxes from 0.3--3.5 to 0.5--2.0 keV band.
We have verified that the same XLF data points of H92 are obtained
within 10\% in the original 0.3--3.5 keV band. We note the excellent
agreement between the RDCS and the EMSS samples in the lowest redshift
bin, where the respective XLFs overlap. More importantly, the EMSS XLF
in the high redshift shell is consistent within the errors with
the RDCS XLF over the luminosity range $8\times 10^{43} - 3\times
10^{44}$. The RDCS however, cannot
follow the drop of the EMSS at the bright end of the XLF.

\section{Cluster Number Counts}

In addition to 
the spectroscopically confirmed sample, a larger number of cluster
candidates have been identified from imaging data. We can now compute 
the cluster number counts from a sample of 130 candidates drawn from 
Sample A+B down $f_{-14}=2$, 
90\% of which have positive identification. We do not attempt here to extend
this analysis below $f_{-14}=2$ since the optical identification is still largely incomplete at such low fluxes. 
The resulting cluster $\log N$--$\log S$,
 which significantly improves the one of R95, is shown in fig.4 along with
previous determinations. 
The error bars take into account the statistical uncertainties as 
discussed in R95 and the systematics due to flux measurements. The latter
typically increase the upper error bar of $N(>\!S)$ by +15\% at high fluxes 
as a result of the possible lost flux in the evaluation of the wavelet algorithm. The faintest three data points have been corrected for the
small 10\% incompleteness. These surface densities are
in good agreement with those of the WARPS survey (Jones et al., 1997)
above their flux limit $f_{-14}=6$.
A maximum likelihood fit to the differential
 number counts with a double power law model,
$K_1 S^{-\alpha_1} $ for $ s\ge S_B$, 
$K_1 S_B^{(\alpha_2-\alpha_1)} S^{-\alpha_2} $ for $ s\le S_B $,
yields: $\alpha_1 = 2.22\pm 0.15$, $\alpha_2 = 1.91\pm 0.18$,
$S_B=1.03\pm 0.05$, $K_1 = 1.10\pm 0.15$ deg$^{-2}$, where fluxes are in
units of $10^{-13}\fun$.
This curve can be integrated to provide the cluster contribution to the soft X-ray background,
$I_{CXB}(0.5-2\, \hbox{keV}) \simeq 1.5 \,\hbox{keV}\, \hbox{cm}^{-2} 
\hbox{s}^{-1} \hbox{sr}^{-1}$  down to $f_{-14}=1$, and is found to 
account for about 10\% of the observed soft extragalactic X-ray background.
 
Given the accurate determination of the local XLF (fig.2), which we 
parametrize as the BCS Schechter XLF with the RDCS faint slope, 
we can now
compare our cluster $\log N$--$\log S$ with the predicted number counts
for a non-evolving XLF. In fig.4 we show that these predictions,
obtained by integrating the local XLF over  the probed redshift
and luminosity range for two values of $q_0$,
are  very close to the observed counts. We note that these number
counts are drawn from a sample which is a factor 2 deeper than the one
used to compute the XLF. The $\log N$--$\log S$ at low fluxes is very
sensitive to the faint end slope of the XLF, as well as its evolution,
which is not directly probed by the RDCS XLF in the two highest
redshift bins. 
A significant steepening (or flattening) ($|\Delta\alpha| >0.2$) of the
faint end ($L_X < 10^{43}\lun$) of the XLF at $z>0.3$ would lead to an
overprediction (or underprediction) of the observed faint number counts
outside the error bars, for any value of $0\le q_0\le 0.5$. This
further strengthens the evidence that the XLF does not evolve, within
the present uncertainties, over a wide range of luminosities ($2\times
10^{42} \lesssim L_X(\lun) \lesssim 3\times 10^{44}$).  The
implications that these faint number counts have for CDM models of
cluster formation have been explored  by \cite{KS97} and \cite{ME97}.

\section{Conclusions}

The large size of the RDCS sample and its wide redshift baseline
 allow us to provide an accurate assessment of the evolution of the
space density of galaxy clusters over a considerable range in X-ray
luminosities and out to the highest redshifts probed so far in X-ray
surveys. We find an excellent agreement with independent determinations 
of the local XLF.
By combining the RDCS XLF with an improved determination of faint
cluster number counts, we find that the XLF does not evolve
significantly at $z\lesssim 0.8$ in the luminosity range ($2\times
10^{42}-3\times 10^{44}\lun$).  Our results extend and complement those
of the EMSS, further restricting the possibility of negative evolution
in the X-ray cluster population at $z\gtrsim 0.3$ to only those clusters
with X-ray luminosities in excess of the local $L^{\displaystyle
\ast}_{[0.5-2] \rm{\footnotesize{keV}}}\simeq 4\times 10^{44} \lun$.

\acknowledgments
We thank the TACs of Kitt Peak National Observatoty, Cerro Tololo
Inter-American Observatory and the European Southern Observatory for
the allocation of generous observing time and the staff of all the
observatories for their excellent support. PR acknowledges support from
NASA ADP grant NAG 5--3537.



\clearpage

\begin{figure}
\plotone{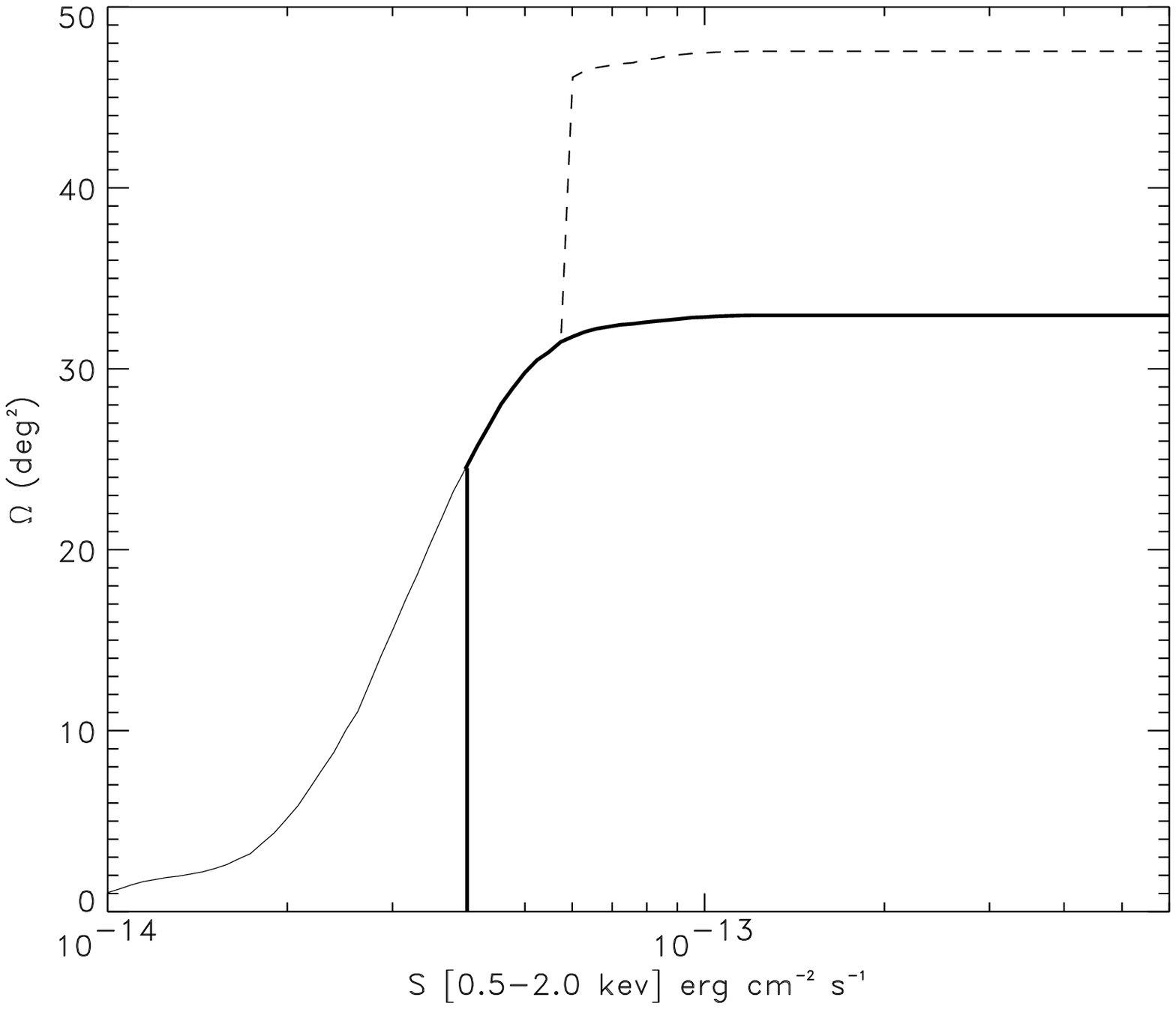}
\caption{The effective sky coverage of the RDCS for Sample A (solid
line) and its extension, Sample A + Sample B (dashed line). The thick solid
line denotes the spectroscopically confirmed subsample used for
computing the XLF.}
\end{figure}

\clearpage
\begin{figure}
\plotone{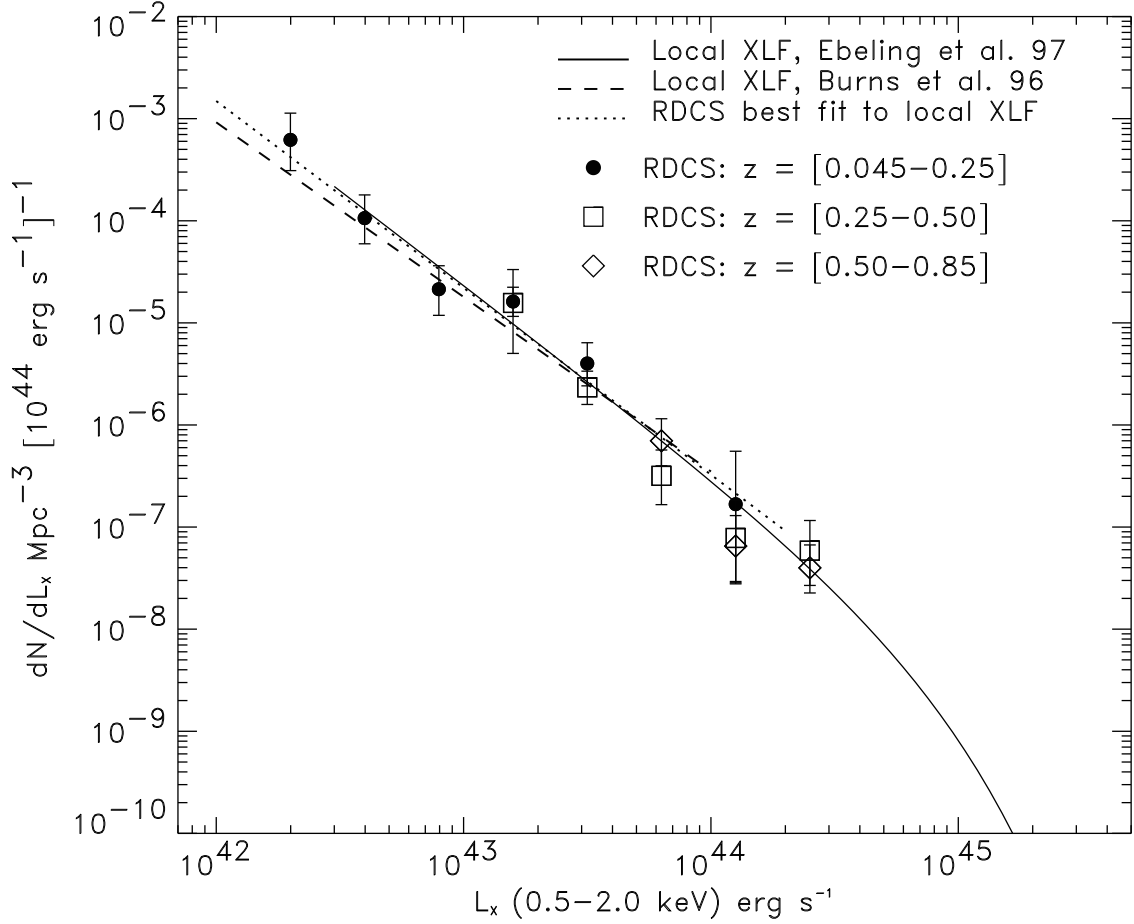}
\caption{The X-ray cluster luminosity function for the RDCS sample  in three redshift shells. Independent determinations of the local XLF from ROSAT All-Sky Survey data are also shown (dashed and solid lines).}
\end{figure}

\clearpage

\begin{figure}
\plotone{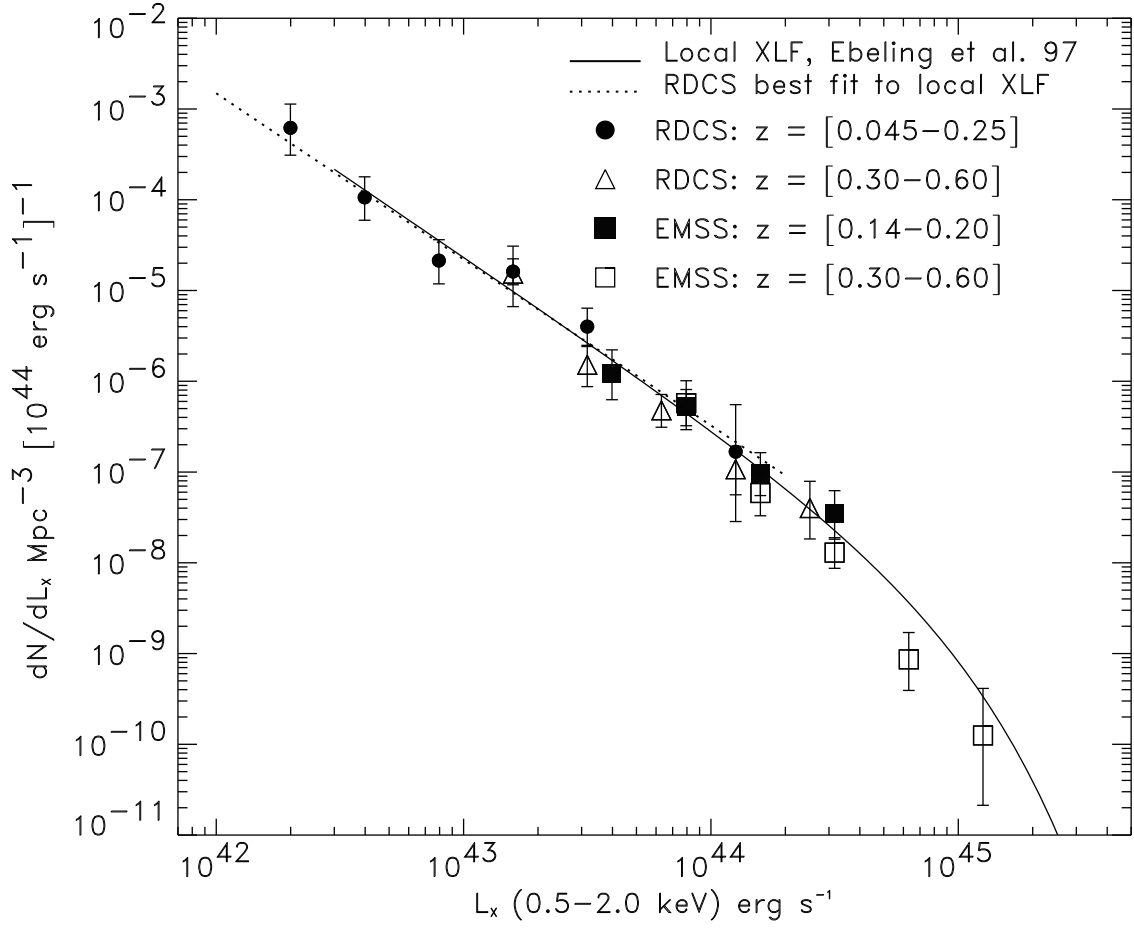}
\caption{Comparison of the luminosty functions for the RDCS and 
the EMSS samples at low and high redshifts.}
\end{figure}

\clearpage

\begin{figure}
\plotone{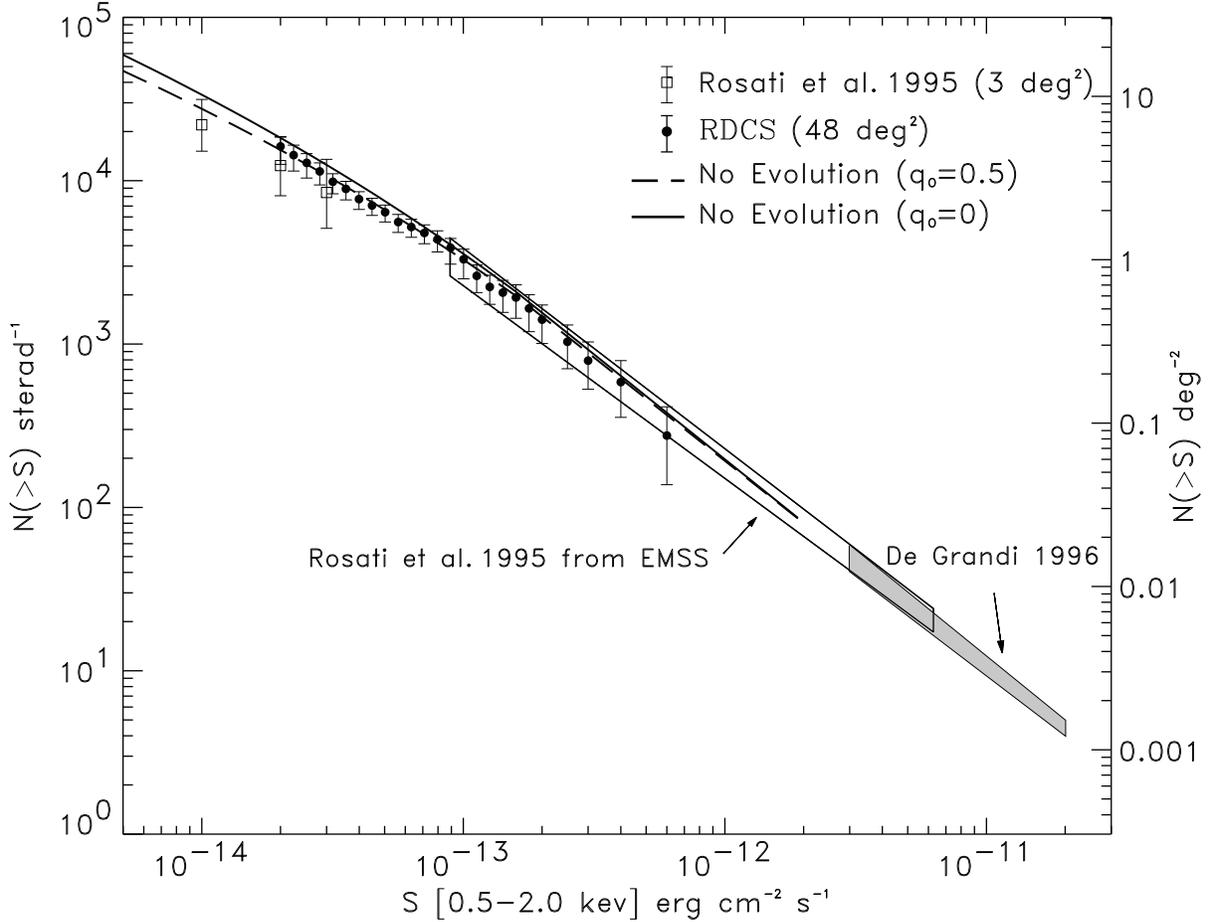}
\caption{Observed cluster cumulative number counts for the RDCS sample
(Sample A + B) and previous determinations.  The no evolution curves
were computed by integrating the best determination of the local XLF
(see text) over the luminosity range $1\times 10^{42}-5\times 10^{45}
\lun$ out to $z=1.1$.}
\end{figure}


\clearpage


\begin{thebibliography}{DUM}

\bibitem[Burke et al. 1997]{bur97} Burke, D.J., Collins, C.A., Sharples,
R.M., Romer, A.K., Holden, B.P., Nicol, R.C. 1997 \apjl, in press
\bibitem[Burns et al. 1996]{B96} Burns, J.O. et al. 1996, \apjl, 467, L49
\bibitem[Castander et al. 1995]{cas95} Castander, F.J. et al. 1995, \nat, 
 377, 39
\bibitem[Collins et al. 1997]{col97} Collins, C.A., Burke, D.J., Romer, A.K.,
 Sharples, R.M., Nichol, R.C. 1997, \apjl, 479, L117
\bibitem[De Grandi, 1996]{sab96} De Grandi, S. 1996, in MPE Report 263, Proceedings of R\"{o}ntgenstrahlung from the Universe, ed. Zimmermann H.U., Tr\"{u}mper, J., Yorke, H. (Munich:MPE), 577 
\bibitem[Ebeling et al. 1997]{ebe97} 
 Ebeling H., Edge, A.C, Fabian, A.C., Allen, S.W., Crawford, C.S.,
 B\"{o}hringer, H. 1997, \apjl, 479, L101
\bibitem[Edge et al. 1990]{edg90} Edge, A.C., Stewart, G.C., Fabian, A.C,
 Arnaud, K.A. 1990, \mnras, 245, 559
\bibitem[Gioia et al. 1990]{gio90} 
 Gioia, I.M., Henry, J.P., Maccacaro, T., Morris, S.L., Stocke, J.T., Wolter, A. 1990, \apjl, 356, L35
\bibitem[Gioia \& Luppino 1994] {giolup94}
 Gioia I.M. \& Luppino, G.A. 1994, \apjs, 94, 583
\bibitem[Henry et al. 1992]{hen92}  
 Henry, J.P., Gioia, I.M., Maccacaro, T.,
 Morris, S.L., Stocke, J.T., Wolter, A. 1992, \apj, 386, 408 (H92)
\bibitem[Jones et al. 1997]{jon97} Jones, L.R., Scharf, C.A., Ebeling, H.,
 Perlman, E., Wegner, G., Malkan, M., Horner, D. 1997, \apj, in press
\bibitem[Kitayama \& Suto (1997)]{KS97} Kitayama, T. \& Suto, S. 1997, \apj, in press
\bibitem[Maccacaro et al. 1991]{mac91} 
  Maccacaro, T., Della Ceca, R., Gioia,
  I.M., Morris, S.L., Stocke, J.T., Wolter, A. 1991, \apj, 374, 117 
\bibitem[Mathiesen \& Evrard (1997)]{ME97} Mathiesen, B. \& Evrard, A.E. 1997, 
 \mnras, in press 
\bibitem[Nichol et al. 1997]{nic97} Nichol, R.C., Holden, B.P., Romer, A.K.,
 Ulmer, M.P., Burke, D.J., Collins, C.A. 1997, \apj, 481, 644
\bibitem[Rosati \& Della Ceca, 1996]{rdc96} Rosati, P. \& Della Ceca, R. 1996, in MPE Report 263, Proceedings of R\"{o}ntgenstrahlung from the Universe, ed. Zimmermann H.U., Tr\"{u}mper, J., Yorke, H. (Munich:MPE), 613 
\bibitem[Rosati et al. 1995]{R95}  
  Rosati, P., Della Ceca, R., Burg, R., 
  Norman, C., Giacconi, R. 1995,  \apjl, 445, L11 (R95) 
\bibitem[Scharf et al. 1997]{sha97} Scharf, C.A., Jones, L.R., Ebeling H.,
 Perlman, E., Malkan, M., Wegner, G. 1997, \apj, 477, 79
\bibitem[Wolter et al. 1994]{wol94}
 Wolter, A., Caccianiga, A., Della Ceca R., Maccacaro, T. 1994, \apj, 433, 29 
\end{thebibliography}
\end{document}